\documentclass[preprint,twocolumn,final,tightenlines,showpacs,notitlepage,10pt,showkeys,unsortedaddress,pra,a4paper,nobibnotes,nofootinbib]{revtex4}
\usepackage{amssymb}

\input{tcilatex}

\begin{document}

\title{Short-time decoherence of Josephson charge qubits}
\author{Xian-Ting Liang\thanks{%
Email address: xtliang@ustc.edu}}
\affiliation{Department of Physics and Institute of Modern Physics, Ningbo University,
Ningbo 315211, China}

\begin{abstract}
In this paper we investigate the short-time decoherence of single Josephson
charge qubit (JCQ). The measure of decoherence is chosen as the maximum norm
of the deviation density operator. It is shown that when the temperature low
enough (for example $T=30mK$), within the elementary gate-operation time $%
\tau ^{g}\sim 12.7ps$, the decoherence is smaller than $10^{-4}$ at present
setup of JCQ. The Josephson charge qubit is suitable to take the blocks for
quantum computations according to the DiVincenzo low decoherence criterion.
\end{abstract}

\maketitle

\section{Introduction}

David DiVincenzo put forward five criteria for the candidates of quantum
computing hardware to be satisfied \cite{DiVincenzo}. One of which is the
low decoherence. An approximate benchmark of the criterion is a fidelity
loss no more than $\sim 10^{-4}$ per elementary quantum gate operation. The
Josephson junction is considered to be a promising physical realization of
qubit. So it is a very significative work to investigate the decoherence of
the qubits based on the Josephson junctions. To perform quantitative
studying of the decoherence for a given system, in general, one must firstly
solve the quantum dynamics problem of the system coupled to the environment.
Many kinds of methods have been used for this purposes \cite{decoherence01} 
\cite{decoherence02}. The decoherence of the qubit based on Josephson
junctions have been studied by many authors. In their investigations the
dynamics appeal to the Redfield formalism \cite{Redfield} or Bloch-type \cite%
{RMP2001} master equations, where the Markov approximation are used \cite%
{Markov}. The approximation is usually used to evaluate approach to the
thermal state at large times. However, a quantum gate operation is finished
in a instantaneous time so the approximation may be not suitable to
investigate the decoherence of the qubit gates for quantum computing
purposes. Recently, V. Privman \emph{et al}. introduced two measures to
quantify the short-time decoherence \cite{Privman} \cite{Fedichkin et al}.
The measures are based on the short-time approximation of the split-operator
and derived from a operator norm $\left\Vert A\right\Vert $ (seeing the
following definition). By using these measures V. Privman \emph{et al}.
investigated some spin-boson models. In this paper we use one of the
measures investigating the short-time decoherence of the single Josephson
charge qubit (JCQ) operations.

\section{JCQ Hamiltonian and a measure of decoherence}

In this section we firstly review the JCQ model and then introduce the
measure of decoherence used in this paper. The single JCQ Hamiltonian is 
\cite{RMP2001}%
\begin{equation}
H_{R}=E_{ch}\left( n-n_{g}\right) ^{2}-E_{J}\cos \varphi .  \label{eq1}
\end{equation}%
Here, $E_{ch}=e^{2}/\left( C_{g}+C_{J}\right) $ is the charging energy; $%
E_{J}=I_{c}\hbar /2e$ is the Josephson coupling energy \cite{SM25-915-1999},
where $I_{c}$ is the critical current of the Josephson junction, $\hbar $
the Planck's constant divided $2\pi ,$ and $e$ the charge of electron; $%
n_{g}=C_{g}V_{g}/2e$ is the dimensionless gate charge, where $C_{g}$ is the
gate capacitance, $V_{g}$ the controllable gate voltage. The number operator 
$n$ of (excess) Cooper pair on the island, and the phase $\varphi $ of the
superconducting order parameter, are quantum mechanically conjugate$.$
Because the Josephson coupling energy $E_{J}$ is much smaller than the
charging energy $E_{ch}$, and both of them are much smaller than the
superconducting energy gap $\Delta $, the Hamiltonian Eq.(\ref{eq1}) can be
parameterized by the number of the Cooper pairs $n$ on the island as \cite%
{RMP2001} \cite{PRL1997}%
\begin{eqnarray}
H_{R} &=&\sum_{n}\left\{ E_{ch}\left( n-n_{g}\right) ^{2}\left\vert
n\right\rangle \left\langle n\right\vert \right.   \nonumber \\
&&\left. -\frac{1}{2}E_{J}\left[ \left\vert n\right\rangle \left\langle
n+1\right\vert +\left\vert n+1\right\rangle \left\langle n\right\vert \right]
\right\} .  \label{eq2}
\end{eqnarray}%
When $n_{g}$ is modulated to a half-integer, say $n_{g}=1/2$ the levels of
two adjacent states are close to each other, the Josephson tunneling mixes
them strongly. When the temperature $T$ is low enough the system can be
reduced to a two-state system (qubit) because all other charge states have
much higher energy and can be neglected. So the Hamiltonian of the system
can \emph{approximately} reads%
\begin{equation}
H_{s}=-\frac{1}{2}B_{z}\sigma _{z}-\frac{1}{2}B_{x}\sigma _{x},  \label{eq3}
\end{equation}%
where $B_{z}=E_{ch}\left( 1-2n_{g}\right) $ and $B_{x}=E_{J}.$ Eq.(\ref{eq3}%
) is similar to the ideal single qubit model \cite{RMP2001}, but it can be
modulated only by changing one parameter $B_{z}.$ However, changing the
parameter $B_{z}$ (through switching the gate voltage) one can perform the
required one-bit operations. If, for example, one chooses the idle state far
to the left from the degeneracy point, the eigenstate lose to $\left\vert
0\right\rangle $ and $\left\vert 1\right\rangle .$ Then switching the system
suddenly to the degeneracy point for a time $\tau $ and back produces a
rotation in spin space \cite{RMP2001},%
\begin{equation}
U_{J}=\exp \left( \frac{iE_{J}\tau }{2}\sigma _{x}\right) =\left( 
\begin{tabular}{ll}
cos$\frac{\tau E_{J}}{2}$ & $i$sin$\frac{\tau E_{J}}{2}$ \\ 
$i$sin$\frac{\tau E_{J}}{2}$ & cos$\frac{\tau E_{J}}{2}$%
\end{tabular}%
\right) .  \label{eq4}
\end{equation}%
This can be obtained by modulating the gate voltage and making $%
B_{z}=E_{ch}\left( 1-2n_{g}\right) =0.$ So it is interesting for our to
investigate the decoherence of this kind of qubit. If we consider the
interaction of the qubit and its environment the total Hamiltonian becomes%
\begin{equation}
H=H_{s}+H_{I}+H_{B},  \label{eq5}
\end{equation}%
where $H_{B}$ is the environment Hamiltonian which is usually modelled by a
bath of an infinite number of the harmonic oscillator models which is
equivalence to an infinity-mode electromagnetic field. The dissipation and
the decoherence of the quantum systems is considered because of the coupling
of the quantum system and the fluctuating electromagnetic field. In the JCQ
model, the coupling of the qubit with the electromagnetic fluctuations can
be modeled by a impedance $Z\left( \omega \right) ,$ placed in series with
the voltage (see Fig.1 of Ref.\cite{chemphys295}). The bath Hamiltonian $%
H_{B}$ and the interaction Hamiltonian $H_{I}$ are 
\begin{equation}
H_{B}=\sum_{k}M_{k},\text{ \ }H_{I}=\Lambda _{s}\sum_{k}N_{k},  \label{eq6}
\end{equation}%
where $\Lambda _{s}=\sigma _{z}$ and%
\begin{equation}
M_{k}=\omega _{k}a_{k}^{\dagger }a_{k},\text{ }N_{k}=g_{k}a_{k}^{\dagger
}+g_{k}^{\ast }a_{k}.  \label{eq7}
\end{equation}%
Here, $\omega _{k}$ are the bath mode frequencies, $a_{k}^{\dagger }$ and $%
a_{k}$ are bosonic create and annihilation operators. $N_{k}$ are the
freedom of environment and it is direct proportion to the fluctuations of
the voltage of the external circuits \cite{RMP2001}\ \cite{chemphys} and $%
g_{k}$ are the coupling constants.

Because the bath modes are infinite, their frequencies can be taken
continuous. The spectral density of the continuous bath modes is%
\begin{equation}
J_{v}\left( \omega \right) =\sum_{k}g_{k}^{2}\delta \left( \omega -\omega
_{k}\right)   \label{eq8}
\end{equation}%
because it can result in a same Johnson-Nyquist relation which can be
obtained from fluctuation-dissipation theorem on this system. On the other
hand, at low-frequency, the spectral density behavior is%
\begin{equation}
J_{v}\left( \omega \right) =\eta \omega ^{s}\exp \left( \omega /\omega
_{c}\right) ,  \label{eq9}
\end{equation}%
where $\omega _{c}$ is the high-frequency cut-off and $\eta $ is the
dimensionless strength of the dissipation. According to Ref.\cite%
{chemphys295} we know that in the particular interest ohmic case $\left(
s=1\right) $ \cite{RMP59-1-1987}, for the model of JCQ coupling with
electromagnetic fluctuations (see Fig.1 of Ref.\cite{chemphys295}), 
\begin{equation}
\eta =\frac{R}{R_{Q}}\left( \frac{C_{t}}{C_{J}}\right) ^{2},  \label{eq10}
\end{equation}%
where $R_{Q}=\left( 2e\right) ^{2}/h$ is the (superconducting) resistance
quantum. From Eqs.(\ref{eq8}) and (\ref{eq9}) we know that for a ohmic bath%
\begin{equation}
g_{k}^{2}W\left( \omega \right) =\eta \omega \exp \left( \omega /\omega
_{c}\right) ,  \label{eq11}
\end{equation}%
where $W\left( \omega \right) $ is the density of states.

There are many measures to characterize the decoherence. Usually the
environment being assumed to be a large macroscopic the interaction with it
leads to the thermal equilibrium at temperature $T$. In this case, Markovian
type approximations can be used to quantified the decoherent process and it
usually yields the exponential decay of the density matrix elements in the
energy basis of the Hamiltonian $H_{s}.$ In this time scale the measures of
entropy and the first entropy are used for quantifying the decoherence. But
the decoherence of the qubit gate operations cannot be characterized by this
methods because the time of the elementary quantum gate operation is much
shorter than the thermal relaxation time. It has been shown that the norms $%
\left\Vert \sigma \right\Vert _{\lambda }$ is useful for describing the
decoherence of the short-time evolution. Here $\sigma $ is the deviation
operator defined as 
\begin{equation}
\sigma \left( \tau \right) =\rho \left( \tau \right) -\rho ^{i}\left( \tau
\right) ,  \label{eq12}
\end{equation}%
where $\rho \left( \tau \right) $ and $\rho ^{i}\left( \tau \right) $ are
density matrixes of the \textquotedblleft real\textquotedblright\ evolution
(with interaction) and the \textquotedblleft ideal\textquotedblright\ one
(without interaction) of the investigated system. $\left\Vert \sigma
\right\Vert _{\lambda }$ is defined as 
\begin{equation}
\left\Vert \sigma \right\Vert _{\lambda }=\sup_{\varphi \neq 0}\left( \frac{%
\left\langle \varphi \right\vert \sigma \left\vert \varphi \right\rangle }{%
\left\langle \varphi \right. \left\vert \varphi \right\rangle }\right) ^{%
\frac{1}{2}}.  \label{eq13}
\end{equation}%
For a qubit, the norm can be given by%
\begin{equation}
\left\Vert \sigma \right\Vert _{\lambda }=\sqrt{\left\vert \sigma
_{10}\right\vert ^{2}+\left\vert \sigma _{11}\right\vert ^{2}}.  \label{eq14}
\end{equation}%
For a given system, the norm $\left\Vert \sigma \right\Vert _{\lambda }$
increase with time, reflecting the decoherence of the system. However, in
general it is oscillated at the system's internal frequency. Thus, the
decohering effect of the bath is better quantified by the maximal operator
norm, $D\left( t\right) .$ This norm is also defined a measure for
characterizing the short-time decoherence. 
\begin{equation}
D\left( \tau \right) =\sup_{\rho \left( 0\right) }\left( \left\Vert \sigma
\left( \tau ,\rho \left( 0\right) \right) \right\Vert _{\lambda }\right) .
\label{eq15}
\end{equation}%
In the following, we will calculate the $\left\Vert \sigma \right\Vert
_{\lambda }$ then the decoherence $D$ of the JCQ.

\section{Decoherence of JCQ operations}

As shown in above subsection, to calculate the decoherence of the JCQ we
must study its evolution under the interaction of the qubit with its
environment. Suppose the initial state of the system be $R\left( 0\right)
=\rho \left( 0\right) \otimes \Theta ,$ where $\rho \left( 0\right) $ is the
initial state of JCQ and $\Theta $ is the initial state of the environment,
which is the product of the bath modes density matrices\ $\theta _{k}$. In
the initial states, each bath mode $k$ is assumed to be thermalized, namely,%
\begin{equation}
\theta _{k}=\frac{e^{-\beta M_{k}}}{\text{Tr}_{k}\left( e^{-\beta
M_{k}}\right) },  \label{eq16}
\end{equation}%
where $\beta =1/kT$, $k$ is the Boltzmann constant. So the evolution
operator may be%
\begin{equation}
U=e^{-iH\tau /\hslash }=e^{-i\left( H_{s}+H_{I}+H_{B}\right) \tau /\hslash }.
\label{eq17}
\end{equation}%
In the following we set $t=\tau /\hslash .$ Due to non-conservation of $H_{s}
$ in this system, the evolution operator cannot be in a general way
expressed as $e^{-iH_{s}t}e^{-i\left( H_{I}+H_{B}\right) t}.$ But in the
sort-time approximation, the operator can be approximately expressed as \cite%
{split-operator01} \cite{split-operator02}%
\begin{equation}
U=e^{-iH_{s}t/2}e^{-i\left( H_{I}+H_{B}\right) t}e^{-iH_{s}t/2}+o(t^{3}).
\label{eq18}
\end{equation}%
It has been proved that the expression is accurate enough for the time being
short to the characteristic time. So the density matrix elements of the
reduced density matrix $\rho \left( t\right) $ in the basis of operator $%
H_{s}$ can be expressed as%
\begin{eqnarray}
\rho _{mn} &=&\text{Tr}_{B}\left\langle \varphi _{m}\right\vert
e^{-iH_{s}t/2}e^{-i\left( H_{I}+H_{B}\right) t}e^{-iH_{s}t/2}R\left(
0\right)   \nonumber \\
&&e^{iH_{s}t/2}e^{i\left( H_{I}+H_{B}\right) t}e^{iH_{s}t/2}\left\vert
\varphi _{n}\right\rangle .  \label{eq19}
\end{eqnarray}%
By use of the completeness relation $\tsum \left\vert \cdot \right\rangle
\left\langle \cdot \right\vert =1$, we have 
\begin{equation}
e^{\pm jH_{s}t/2}=\tsum_{j=0,1}e^{\pm it\lambda _{j}}\left\vert \varphi
_{j}\right\rangle \left\langle \varphi _{j}\right\vert ,  \label{eq20}
\end{equation}%
where%
\begin{equation}
\lambda _{0,1}=\pm \frac{B_{x}}{2}=\pm \frac{E_{J}}{2},  \label{eq21}
\end{equation}%
and 
\begin{eqnarray}
\left\vert \varphi _{0}\right\rangle  &=&\frac{1}{\sqrt{2}}\left( \left\vert
0\right\rangle -\left\vert 1\right\rangle \right) ,  \nonumber \\
\left\vert \varphi _{1}\right\rangle  &=&\frac{1}{\sqrt{2}}\left( \left\vert
0\right\rangle +\left\vert 1\right\rangle \right) .  \label{eq22}
\end{eqnarray}%
Similarly, 
\begin{equation}
e^{\pm i\left( H_{I}+H_{B}\right) t}=\tsum_{n=0,1}e^{\pm i\left( \chi
_{n}\tsum_{k}J_{k}+\tsum_{k}M_{k}\right) t}\left\vert n\right\rangle
\left\langle n\right\vert ,  \label{eq23}
\end{equation}%
where $\chi _{0,1}=\pm 1$ and $\left\vert n\right\rangle =\left\vert
0\right\rangle $ or $\left\vert 1\right\rangle ,$ are the eigenvalues and
eigenstates of operator $\sigma _{z}.$ So we have%
\begin{eqnarray}
\rho _{mn}\left( t\right)  &=&\text{Tr}_{B}\left\langle \varphi
_{m}\right\vert \tsum_{\alpha =0,1}e^{-it\lambda _{\alpha }}\left\vert
\varphi _{\alpha }\right\rangle \left\langle \varphi _{\alpha }\right\vert  
\nonumber \\
&&\tsum_{\xi =0,1}e^{-i\left( \chi _{\xi }\sum_{k}J_{k}+\sum_{k}M_{k}\right)
t}\left\vert \xi \right\rangle \left\langle \xi \right\vert   \nonumber \\
&&\tsum_{\beta =0,1}e^{-it\lambda _{\beta }}\left\vert \varphi _{\beta
}\right\rangle \left\langle \varphi _{\beta }\right\vert
\tsum_{p,q=0,1}\left\vert \varphi _{p}\right\rangle \left\langle \varphi
_{q}\right\vert   \nonumber \\
&&\rho _{pq}\left( 0\right) \tprod_{k}\theta _{k}\tsum_{\mu
=0,1}e^{it\lambda _{\mu }}\left\vert \varphi _{\mu }\right\rangle
\left\langle \varphi _{\mu }\right\vert   \nonumber \\
&&\tsum_{\varsigma =0,1}e^{i\left( \chi _{\varsigma
}\tsum_{k}J_{k}+\tsum_{k}M_{k}\right) t}\left\vert \varsigma \right\rangle
\left\langle \varsigma \right\vert   \nonumber \\
&&\tsum_{\nu =0,1}e^{it\lambda _{\nu }}\left\vert \varphi _{\nu
}\right\rangle \left\langle \varphi _{\nu }\right\vert \left. \varphi
_{n}\right\rangle ,  \label{eq24}
\end{eqnarray}%
namely,%
\begin{eqnarray}
\rho _{mn}\left( t\right)  &=&\tsum_{\alpha ,\beta ,\xi ,\varsigma ,p,q,\mu
,\nu =0,1}e^{it\left( \lambda _{\mu }+\lambda _{\nu }-\lambda _{\alpha
}-\lambda _{\beta }\right) }\left\langle \varphi _{m}\right\vert \left.
\varphi _{\alpha }\right\rangle   \nonumber \\
&&\left\langle \varphi _{\alpha }\right\vert \left. \xi \right\rangle
\left\langle \xi \right\vert \left. \varphi _{\beta }\right\rangle
\left\langle \varphi _{\beta }\right\vert \left. \varphi _{p}\right\rangle
\left\langle \varphi _{q}\right\vert \left. \varphi _{\mu }\right\rangle
\left\langle \varphi _{\mu }\right\vert \left. \varsigma \right\rangle  
\nonumber \\
&&\left\langle \varsigma \right\vert \left. \varphi _{\nu }\right\rangle
\left\langle \varphi _{\nu }\right\vert \left. \varphi _{n}\right\rangle 
\text{Tr}_{B}\left[ e^{-i\left( \chi _{\xi
}\tsum_{k}J_{k}+\tsum_{k}M_{k}\right) t}\right.   \nonumber \\
&&\left. \rho _{pq}\left( 0\right) \tprod_{k}\theta _{k}e^{i\left( \chi
_{\varsigma }\tsum_{k}J_{k}+\tsum_{k}M_{k}\right) t}\right] .  \label{eq25}
\end{eqnarray}%
In the Eq.(\ref{eq25}) the Tr$_{B}$ term is same as that in Ref.\cite%
{Privman}. Enlightened by Privman's works we can easily obtain%
\begin{eqnarray}
\rho _{mn}\left( t\right)  &=&\tsum_{\alpha ,\beta ,\xi ,\varsigma ,p,q,\mu
,\nu =0,1}e^{it\left( \lambda _{\mu }+\lambda _{\nu }-\lambda _{\alpha
}-\lambda _{\beta }\right) }  \nonumber \\
&&\left\langle \varphi _{m}\right\vert \left. \varphi _{\alpha
}\right\rangle \left\langle \varphi _{\alpha }\right\vert \left. \xi
\right\rangle \left\langle \xi \right\vert \left. \varphi _{\beta
}\right\rangle \left\langle \varphi _{\beta }\right\vert \left. \varphi
_{p}\right\rangle   \nonumber \\
&&\left\langle \varphi _{q}\right\vert \left. \varphi _{\mu }\right\rangle
\left\langle \varphi _{\mu }\right\vert \left. \varsigma \right\rangle
\left\langle \varsigma \right\vert \left. \varphi _{\nu }\right\rangle
\left\langle \varphi _{\nu }\right\vert \left. \varphi _{n}\right\rangle  
\nonumber \\
&&\rho _{pq}\left( 0\right) e^{-B^{2}\left( t\right) \left( \chi _{\xi
}-\chi _{\varsigma }\right) ^{2}/4-iC\left( t\right) \left( \chi _{\xi
}^{2}-\chi _{\varsigma }^{2}\right) }.  \label{eq26}
\end{eqnarray}%
Here, 
\begin{eqnarray}
B^{2}\left( t\right)  &=&8\tsum_{k}\frac{\left\vert g_{k}\right\vert ^{2}}{%
\omega _{k}^{2}}\sin ^{2}\frac{\omega _{k}t}{2}\coth \frac{\beta \omega _{k}%
}{2},  \nonumber \\
C\left( t\right)  &=&\tsum_{k}\frac{\left\vert g_{k}\right\vert ^{2}}{\omega
_{k}^{2}}\left( \omega _{k}t-\sin \omega _{k}t\right) ,  \label{eq27}
\end{eqnarray}%
and $\left\vert \xi \right\rangle ,$ $\left\vert \varsigma \right\rangle \in
\left\{ \left\vert 0\right\rangle ,\left\vert 1\right\rangle \right\} .$ $%
B^{2}\left( t\right) $ will affect the decoherence but $C\left( t\right) $
will not it quantifies purely a shift of the energy levels of the qubit
system. When the summation in Eq.(\ref{eq27}) is converted to integration in
the limit of infinite number of the bath modes, one has%
\begin{equation}
B^{2}\left( t\right) =8\int d\omega W\left( \omega \right) g\left( \omega
\right) ^{2}\omega ^{-2}\sin ^{2}\frac{\omega t}{2}\coth \frac{\beta \omega 
}{2},  \label{eq28}
\end{equation}%
for the real $g\left( \omega \right) .$ By using of Eq.(\ref{eq11}) we can
yield a good qualitative estimate of the relaxation behavior \cite{Privman}, 
\cite{JSP78-299-1995}. From Ref.\cite{RMP59-1-1987} we know the condition of
Eq.(\ref{eq9}) is $\omega _{c}\hbar \gg E_{J}$ and $\omega _{c}\hbar \gg
k_{B}T$. It is big enough for $\omega _{c}$ to take $\omega _{c}\hbar
=200\mu ev$. Thus, in the following numerical simulations we set the cutoff
frequency $\omega _{c}=200$ (the unit is the reciprocal of $t^{\prime }s$).
Evaluating Eq.(\ref{eq26}) we can obtain the evolution of the density matrix
elements of the deduce density matrix $\rho $ as%
\begin{eqnarray}
\rho _{10}\left( t\right)  &=&\frac{1}{2}\rho _{10}\left( 1-e^{-B^{2}\left(
t\right) }+e^{itE_{J}}+e^{itE_{J}-B^{2}\left( t\right) }\right) ,  \nonumber
\\
\rho _{11}\left( t\right)  &=&\frac{1}{2}\rho _{00}\left( 1-e^{-B^{2}\left(
t\right) }\right) +\frac{1}{2}\rho _{11}\left( 1+e^{-B^{2}\left( t\right)
}\right) ,  \label{eq29}
\end{eqnarray}%
where $\rho _{11}=\rho _{11}\left( 0\right) $, $\rho _{10}=\rho _{10}\left(
0\right) .$ The evolution of the closed qubit is $\rho _{11}^{i}\left(
t\right) =\rho _{11},$ and $\rho _{10}^{i}\left( t\right) =\rho
_{10}e^{itE_{J}}.$ So we have%
\begin{eqnarray}
\sigma _{10}\left( t\right)  &=&\frac{1}{2}\rho _{10}\left(
1-e^{-B^{2}\left( t\right) }\right) \left( 1-e^{itE_{J}}\right) ,  \nonumber
\\
\sigma _{11}\left( t\right)  &=&\frac{1}{2}\left( 1-e^{-B^{2}\left( t\right)
}\right) \left( \rho _{00}-\rho _{11}\right) .  \label{eq30}
\end{eqnarray}%
Then, we have%
\begin{eqnarray}
\left\Vert \sigma \left( t\right) \right\Vert _{\lambda } &=&\frac{1}{2}%
\left( 1-e^{-B^{2}\left( t\right) }\right)   \nonumber \\
&&\left\{ \left( \rho _{00}-\rho _{11}\right) ^{2}+4\left\vert \rho
_{10}\right\vert ^{2}\sin ^{2}\frac{E_{J}t}{2}\right\} ^{\frac{1}{2}}.
\label{eq31}
\end{eqnarray}%
The result is similar but not same to the norm in \cite{Privman} where $%
H_{s}=-\frac{\Omega }{2}\sigma _{z}$ and $\Lambda _{s}=\sigma _{x}.$

In the following, we numerically analyze the decoherence. In the following
calculation, three pure initial states are chosen, they are $\left\vert
\varphi \right\rangle _{0}=\left( 1,0\right) ^{T},$ (corresponding to
point); $\left\vert \varphi \right\rangle _{1}=\left( \frac{\sqrt{3}}{2},%
\frac{1}{2}\right) ^{T},$ (corresponding to above line); $\left\vert \varphi
\right\rangle _{2}=\left( \sqrt{\frac{1}{2}},\sqrt{\frac{1}{2}}\right) ^{T},$
(corresponding to below line), [because it has been shown that evaluation of
the supremum over the initial density operators in order to find $D,$ see
Eq.(\ref{eq15}) one can do over only pure-state density operators \cite%
{Privman}]. We choose $E_{J}=51.8\mu ev$ according to \cite{Nakamura}, and $%
T=30mK.$ In the model the typical impedance of the control line is $R\approx
50\Omega .$ Since $R_{Q}\approx 6.5k\Omega ,$ Y. Maklin \emph{et al}. have
suggested a value $\eta \approx 10^{-6}$ for numerical$\ $simulations$.$ We
plot the norms $\left\Vert \sigma \right\Vert _{\lambda }$ versus time $t$
in Fig.1. 
\begin{eqnarray*}
&& \\
&& \\
&&\text{ \ \ \ \ \ \ \ \ \ \ \ \ }Fig1.a,\text{ }Fig1.b \\
&& \\
&& \\
&& \\
&&%
\begin{tabular}{l}
\emph{Fig.1: Norms }$\left\Vert \sigma \right\Vert _{\lambda }$\emph{\
versus time }$t$,\emph{\ where the points}%
\end{tabular}
\\
&&%
\begin{tabular}{l}
\emph{and lines correspond to different initial states (see}%
\end{tabular}
\\
&&%
\begin{tabular}{l}
\emph{the\ text),\ }$E_{J}=51.8\mu ev\emph{,\ }T=30mk$ \emph{and\ }$\eta
=10^{-6}$.%
\end{tabular}
\\
&&%
\begin{tabular}{l}
$\emph{The\ unit\ of\ the\ time\ in\ }$\emph{the\ Figs. is }$6.582\times
10^{-10}s$.%
\end{tabular}%
\end{eqnarray*}%
It is shown that when the initial state is $\rho \left( 0\right) =\left\vert
\varphi \right\rangle _{00}\left\langle \varphi \right\vert ,$ $\left\Vert
\sigma \left( t\right) \right\Vert _{\lambda }$ is the maximum and it equals
to $D=\frac{1}{2}\left( 1-e^{-B^{2}\left( t\right) }\right) $ (plotted by
points in the Figs.). We denote the low decoherence ($D\leq 10^{-4}$) time $%
\tau ^{ld}.$ From Fig.1a we obtain $\tau ^{ld}\approx 7.5\times
10^{-2}\times 6.582\times 10^{-10}s=49.4ps.$ We denote the elementary gate
operation time, the characteristic time $\tau ^{g}.$ In this case, $\tau
^{g}=\hbar /E_{J}\approx 12.7ps$. It is shown that the low decoherence time
is larger than the single gate operation time, namely, $\tau ^{ld}>\tau ^{g}$%
. The fact means that within the whole time of elementary gate operation, $%
D\leq 10^{-4}.$ Theoretically, the design can satisfy the DiVincenzo low
decoherence criterion. A further study shows that when the temperature
decreased and the Josephson energy $E_{J}$ increased the qubit can be
improved. It is also shown that the decoherence increase with the increasing
of dimensionless strength of the dissipation $\eta .$

\section{Conclusions}

In this paper we investigated the short-time decoherence of the JCQ. We show
schematically the behavior of $\left\Vert \sigma \left( t\right) \right\Vert
_{\lambda }$. It is shown that for a JCQ the decoherence derive from the
dissipation is small enough according to the DiVincenzo criterion. It has
been shown that the decoherence of JCQ derive not only from the dissipation
but also from the quantum leakage. Fazio et al. pointed that the decoherence
from quantum leakage is also not serious\ in JCQ model \cite{Fazio}. The two
aspects information shows that the JCQ may be a good candidate of qubit for
quantum computation.

\begin{acknowledgments}
I would like to thank Prof. Vladimir Privman and Dr. Leonid Fedichkin for
their helpful discussions. A comment by an anonymous referee has greatly
improved the paper. The project was supported by National Natural Science
Foundation of China (Grant No. 10347133) and Ningbo Youth Foundation (Grant
No. 2004A620003).
\end{acknowledgments}

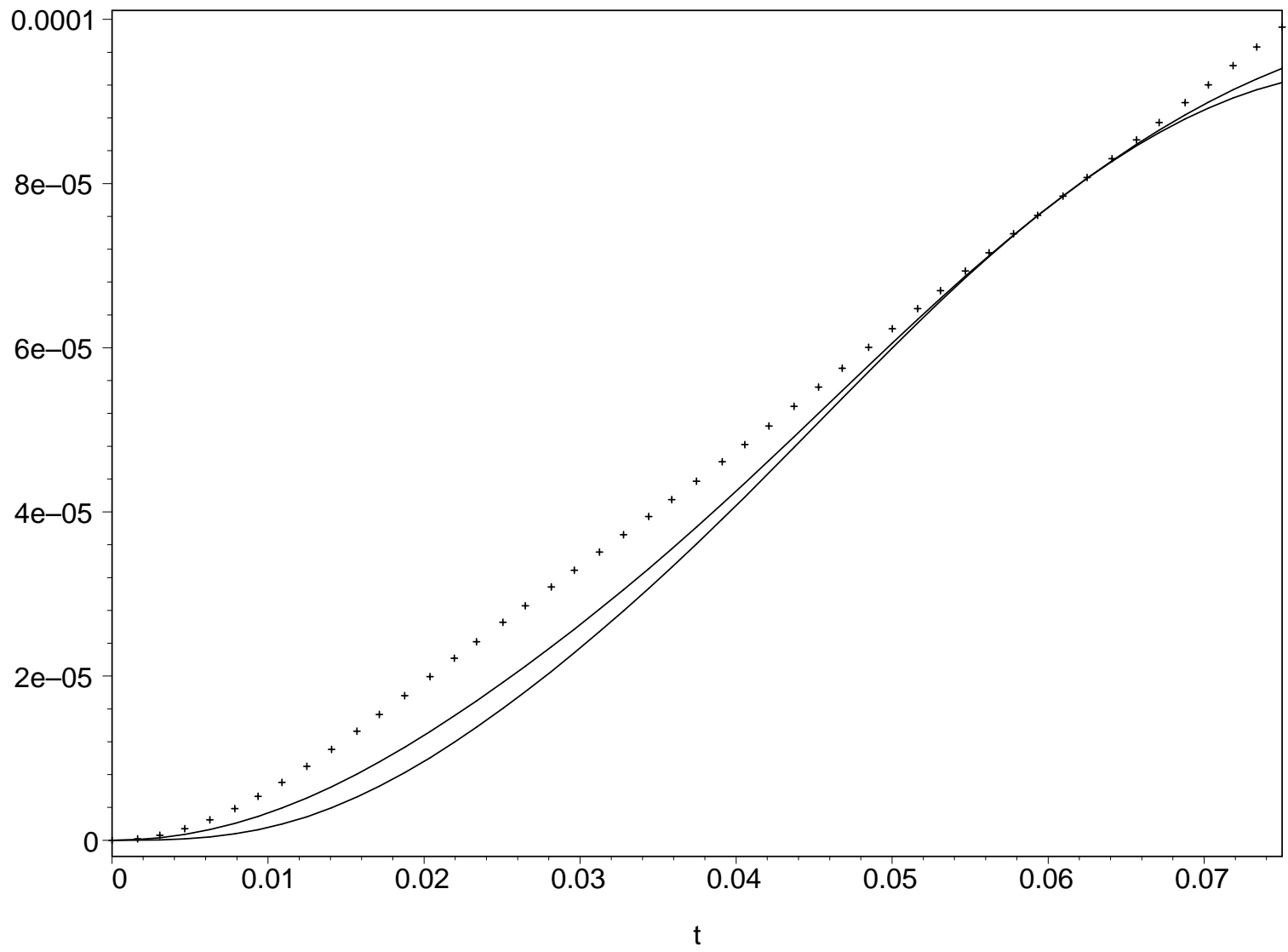
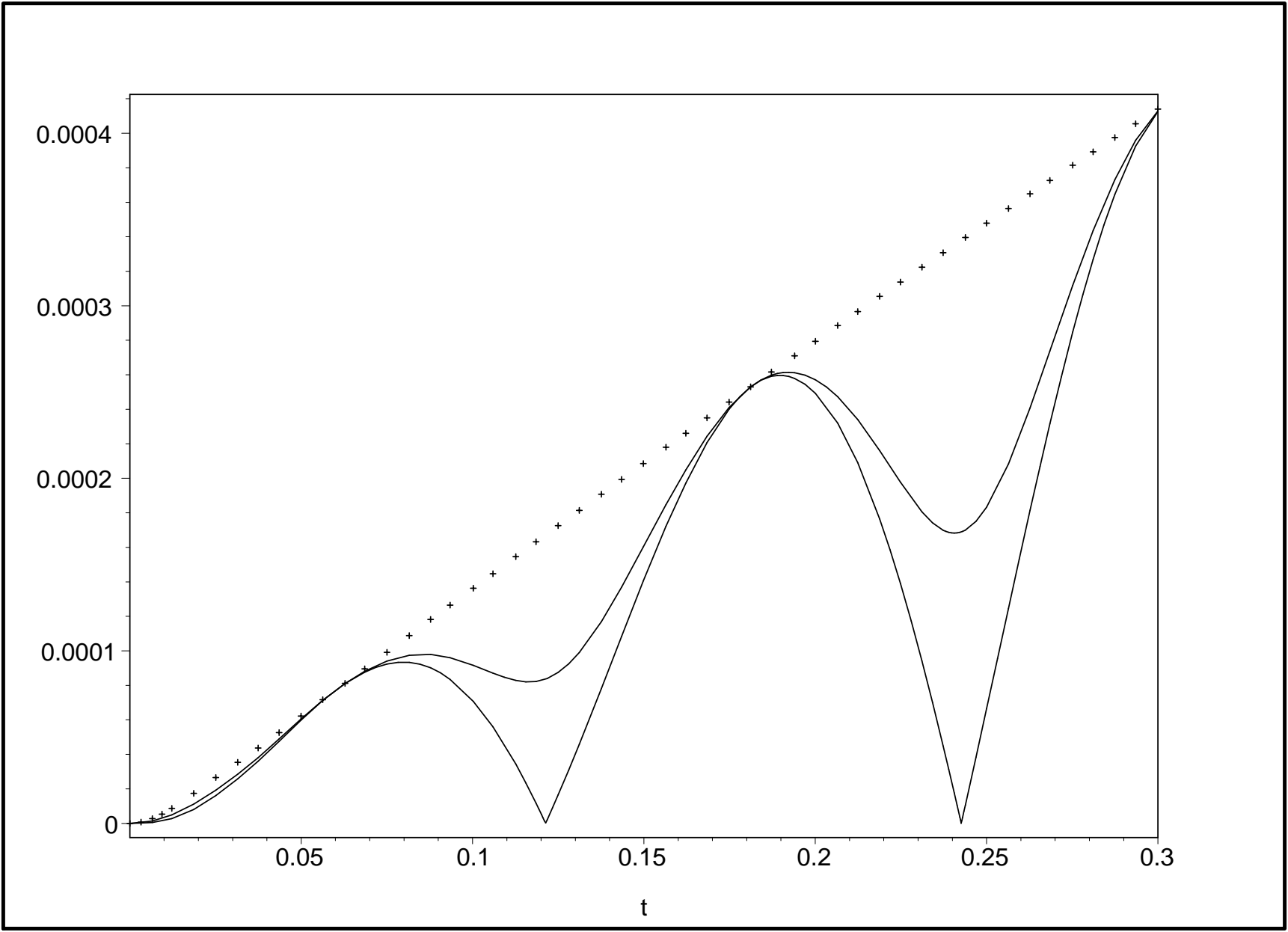
\end{document}